\documentclass[%
reprint,
superscriptaddress,
 amsmath,amssymb,
 aps,
prb,
nobalancelastpage,
]{revtex4-2}

\usepackage{amsmath}
\usepackage{graphicx}
\usepackage{dcolumn}
\usepackage{bm}
\usepackage{hyperref}
\usepackage{url}
\usepackage{xcolor}

\hypersetup{colorlinks=true,allcolors=blue}

\newcommand{\Sr}{\texorpdfstring{Sr$_{2}$RuO$_{4}$}{Sr2RuO4}}

\begin{document}

\preprint{APS/123-QED}

\title{\texorpdfstring{Higher angular momentum pairings in interorbital shadowed-triplet superconductors:\\ Application to \protect\Sr}{Higher angular momentum pairings in interorbital shadowed-triplet superconductors:Application to \protect\Sr}}

\author{Jonathan Clepkens}
\affiliation{Department of Physics and Center for Quantum Materials,\\ University of Toronto, 60 St. George St., Toronto, Ontario, M5S 1A7, Canada}
\author{Austin W. Lindquist}
\affiliation{Department of Physics and Center for Quantum Materials,\\ University of Toronto, 60 St. George St., Toronto, Ontario, M5S 1A7, Canada}
\author{Xiaoyu Liu}
\affiliation{Department of Physics and Center for Quantum Materials,\\ University of Toronto, 60 St. George St., Toronto, Ontario, M5S 1A7, Canada}
\author{Hae-Young Kee}
 \email{hykee@physics.utoronto.ca}
\affiliation{Department of Physics and Center for Quantum Materials,\\ University of Toronto, 60 St. George St., Toronto, Ontario, M5S 1A7, Canada}
\affiliation{Canadian Institute for Advanced Research, Toronto, Ontario, M5G 1Z8, Canada}


\begin{abstract}
Even-parity interorbital spin-triplet pairing emerges as an intriguing candidate in multiorbital superconductors with significant Hund's and spin-orbit coupling. Within such a state, the pairing is dominated by the intraband pseudospin-singlet component via the spin-orbit coupling, distinguishing it from a pure spin triplet and motivating the name, shadowed triplet. With atomic spin-orbit coupling, it was shown that this pairing exhibits an anisotropic $s$-wave character, while higher angular momentum pairing symmetries such as $d$- or $g$-wave have been proposed based on phenomenological analyses in candidate systems. A natural question is then whether higher angular momentum pairings may arise with this form of spin-triplet pairing. Here, we examine the interplay between spin-orbit coupling and the electronic dispersions in correlated metals and demonstrate how they can be realized. We apply this idea to Sr$_{2}$RuO$_{4}$ and determine the competition among different pairing states as multiple spin-orbit coupling parameters are tuned. The presence of both $d$- and $g$-wave pairings, including a $d+ig$ state, are found when momentum-dependent spin-orbit coupling with $d$-wave character is increased. Implications of the theory and future directions are also discussed.
\end{abstract}

\maketitle

\section{\label{intro}Introduction}

In multiorbital materials where several orbitals are relevant for the low-energy electronic states, exotic forms of unconventional superconductivity are possible due to the increased degrees of freedom. For instance, proposals ranging from orbitally selective pairing states to interorbital pairing have been explored in iron-based and heavy-fermion superconductors \cite{KarbowskiPRB1994,MasudaBookJPS,ZhouPRB2008,Ongpnas2016,GaoPRB2010,nica2017nature,nica2021multiorbital}, as well as for the prominent ruthenate-based superconductor, Sr$_{2}$RuO$_{4}$ \cite{Maeno1994Nature,Mackenzie2003RMP,Kallin2012rpp,Mackenzie2017NPJ, SpalekPRB2001,Puetter2012EPL, Hoshino2015PRL, Hoshino2016PRB,Gingras2019PRL, Suh2019, clepkens2021, kaser2021}. In particular, beyond even-parity spin-singlet and odd-parity spin-triplet pairings, multiorbital systems allow for even-parity spin-triplet or odd-parity spin-singlet pairings when the Cooper pair wave function is antisymmetric in the orbital index. One mechanism for this is provided by the Hund's coupling, which allows for a local spin-triplet pair to form between electrons in different orbitals, i.e., even-parity interorbital spin-triplet pairing \cite{klejnberg1999hund, SpalekPRB2001, HanPRB2004, Dai2008PRL,Puetter2012EPL,Hoshino2015PRL,Hoshino2016PRB, Gingras2019PRL,vafek2017hund}. Indeed, Hund's coupling has been recognized to be crucial for understanding the properties of multiorbital metals such as Sr$_{2}$RuO$_{4}$ \cite{CuocoPRB1998,mravlje2011PRL,LucaPRL2011,Georges2013ARCMP, Strand2019PRB,Tamai2019PRX}.

However, interorbital pairing is fragile as it requires orbital degeneracy, i.e., the energies of both orbitals at the two momenta associated with the Cooper pair, \textbf{k} and $-$\textbf{k}, need to be close to the Fermi surface (FS). The lack of orbital degeneracy at the FS caused by different orbital dispersions prevents this pairing momentum phase space, and thus suppresses the pairing. Other kinetic terms, such as interorbital hopping, can further weaken the interorbital spin-triplet pairing by shifting the bands being paired apart in energy, without providing the required spin-orbital mixing for intraband pairing to occur \cite{Ramires2016PRB,Ramires2018PRB,Cheung2019PRB,clepkens2021}.


In light of the fragility of interorbital pairing, it was recognized that spin-orbit coupling (SOC) is critical for stabilizing the pairing state, by mixing the spin and orbital degrees of freedom within the bands \cite{Puetter2012EPL, vafek2017hund, Cheung2019PRB, Suh2019}. In this case, Cooper pairs are no longer well-defined in terms of spin singlet and triplet, yet with both inversion and time-reversal symmetries, the symmetry-protected twofold degeneracy at each momentum allows for a pseudospin classification of the pairing. If the SOC is sufficiently large, such that the low-energy electronic states are well described by the total angular momentum, the pairing can be classified as either pseudospin singlet or triplet, as in heavy-fermion materials \cite{Joynt2002RMP, Kallin2016}. 

However, when the SOC is comparable to the orbital degeneracy splitting terms, such that the spin and orbital character varies over the FS, like in Sr$_{2}$RuO$_{4}$, the pseudospin character is generally \textbf{k} dependent \cite{Puetter2012EPL,Veenstra2014PRL}. This is reflected by the transformation from interorbital spin-triplet pairing in the orbital basis to both \textbf{k}-dependent pseudospin-singlet and -triplet pairings in the band basis through SOC. The resulting intraband pseudospin-singlet pairing component stabilizes the pairing state and dominates the low-energy response which is similar to a pure singlet \cite{Yu2018PRB,lindquist2019distinct}. The underlying triplet nature is apparent at larger energy scales, for example, under a magnetic field \cite{lindquist2019distinct}, which motivated the name ``shadowed triplet" \cite{clepkens2021}.

Since the pairing interaction provided by the Hund's coupling is local, within this scenario it leads to anisotropic $s$-wave pairings. Here, the anisotropic \textbf{k} dependence originates from the atomic SOC, as discussed above. However, momentum-dependent SOC (\textbf{k}-SOC) can help stabilize pairing states with nontrivial symmetry, encoded by the spin and orbital degrees of freedom. The symmetry of the pairing state is then determined by that of the SOC \cite{Cheung2019PRB, Suh2019, clepkens2021}. For the case of a tetragonal lattice with point group $D_{4h}$, there are various \textbf{k}-SOCs that are allowed by symmetry \cite{Ramires2019PRB}, which can be derived through oxygen-mediated hopping \cite{clepkens2021}. However, the \textbf{k}-SOC parameters are likely small, and a natural question is whether it is possible to have higher angular momentum pairing without the corresponding \textbf{k}-SOC.

Here, we consider the possibility of higher angular momentum pairings such as $d$- or $g$-wave, without necessarily requiring \textbf{k}-SOC with the same $d$- or $g$-wave symmetry. We apply this idea to Sr$_{2}$RuO$_{4}$, for which SOC has been acknowledged to play an important role \cite{Ng2000EPL,EreminPRB2002,AnnettPRB2006, Pavarini2006PRB,Haverkort2008PRL,IwasawaPRL2010,Rozbicki2011JPCM,Veenstra2014PRL,Kim2018PRL,Tamai2019PRX}, to investigate the possibility of $d_{x^2-y^2}$- and $g_{xy(x^2-y^2)}$-wave pairings. We show that the interplay between SOC and the electronic orbital dispersions is crucial to obtaining such higher angular momentum pairing states.

The paper is organized as follows. We first discuss the general microscopic model used throughout, including the kinetic Hamiltonian and the Hubbard-Kanamori interaction terms. We use mean-field (MF) theory for the spin-triplet and -singlet channels in Sec.~\hyperref[Two]{II}. We begin with the case of two orbitals to review the importance of SOC in determining the pairing symmetry. We then consider the case of three $t_{2g}$ orbitals and show that the combination of SOC and other kinetic terms generates pairing states beyond that of the two-orbital case in Sec.~\hyperref[Three]{III}. In Sec.~\hyperref[Four]{IV}, we examine the application to Sr$_{2}$RuO$_{4}$ using a tight-binding (TB) parameter set obtained by first-principles calculations, and subsequently discuss the phase diagram with several competing higher angular momentum pairing states, which are close in energy for a subset of the SOC parameter space. We summarize the results and discuss possible implications and open questions in the last section.

\section{\label{Two}Microscopic Hamiltonian}

We consider a general model with inversion and time-reversal symmetries containing multiple orbitals with the Hamiltonian:
\begin{equation}
    H = H_{0} + H_{\text{SOC}} + H_{\text{int}}.
\end{equation}
The kinetic term, $H_{0}$, denotes a TB model which includes the orbital dispersions and interorbital hoppings,
\begin{equation}
\label{TBgeneral}
    H_{0} = \sum_{\textbf{k}\sigma a} \xi^{a}_{\textbf{k}}c_{a,\textbf{k}\sigma}^{\dagger}c_{a,\textbf{k}\sigma} + \sum_{\textbf{k}\sigma, a\neq b} t_{\textbf{k}}^{a/b}c_{a,\textbf{k}\sigma}^{\dagger}c_{b,\textbf{k}\sigma},
\end{equation}
where $c_{a,\textbf{k}\sigma}$ annihilates an electron in orbital $a$, with spin $\sigma$ and wave-vector $\textbf{k}$, and precise forms for the dispersions ($\xi_{\textbf{k}}^{a}$) and interorbital hoppings ($t_{\textbf{k}}^{a/b}$) will be specified later, along with the orbitals. The SOC, $H_{\text{SOC}}$, generally includes both on-site atomic and even-parity \textbf{k}-SOC, the form of which will also be specified later. Finally, we include the on-site Hubbard-Kanamori interactions given by
\begin{equation}
\begin{aligned}
H_{\text{int}} = &\frac{U}{2}\sum_{i,a,\sigma\neq\sigma'}n_{a,i\sigma}n_{a,i\sigma'} + \frac{U'}{2}\sum_{i,a\neq b,\sigma\sigma'}n_{a,i\sigma}n_{b,i\sigma'}\\& + \frac{J_{H}}{2}\sum_{i, a\neq b,\sigma\sigma'}c^{\dagger}_{a,i\sigma}c^{\dagger}_{b,i\sigma'}c_{a,i\sigma'}c_{b,i\sigma} \\&+\frac{J_{H}}{2}\sum_{i, a\neq b,\sigma\neq\sigma'}c^{\dagger}_{a,i\sigma}c^{\dagger}_{a,i\sigma'}c_{b,i\sigma'}c_{b,i\sigma},
\end{aligned}
\end{equation}
where $U$ and $U'$ are the intraorbital and interorbital Hubbard repulsions, respectively, $J_{H}$ is the Hund's coupling, and $i$ is the site index. The interaction Hamiltonian can be written in terms of even-parity spin-singlet and -triplet pairing channels \cite{Puetter2012EPL,vafek2017hund,Suh2019,lindquist2019distinct,clepkens2021} as
\begin{equation} \label{interactions}
\begin{aligned}
&H_{\text{int}} = \frac{4U}{N}\sum_{a,\textbf{k}\textbf{k}'}\hat{\Delta}^{s\dagger}_{a,\textbf{k}}\hat{\Delta}^{s}_{a,\textbf{k}'} \\&+ \frac{2(U'-J_{H})}{N}\sum_{\{a\neq b\},\textbf{k}\textbf{k}'}\hat{\mathbf{d}}_{a/b,\textbf{k}}^{\dagger}\cdot\hat{\mathbf{d}}_{a/b,\textbf{k}'} \\&+\frac{4J_{H}}{N}\sum_{a\neq b,\textbf{k}\textbf{k}'}\hat{\Delta}^{s\dagger}_{a,\textbf{k}}\hat{\Delta}^{s}_{b,\textbf{k}'}\\&+ \frac{2(U'+J_{H})}{N}\sum_{\{a\neq b\},\textbf{k}\textbf{k}'}\hat{\Delta}^{s\dagger}_{a/b,\textbf{k}}\hat{\Delta}^{s}_{a/b,\textbf{k}'},
\end{aligned}
\end{equation}
where $N$ is the number of sites and $\{a\neq b\}$ represents a sum over the unique pairs of orbital indices. The pairing operators are local in the orbital basis, i.e., they contain no momentum dependence, yet the momentum dependence of the gap becomes apparent when written in the band basis, as shown later. They are defined as
\begin{equation} \label{OPs}
\begin{aligned}
\hat{\textbf{d}}_{a/b,\textbf{k}} &= \frac{1}{4}\sum_{\sigma\sigma'}[i\sigma_{2}\boldsymbol{\sigma}]_{\sigma\sigma'}\bigl(c_{a,-\textbf{k}\sigma}c_{b,\textbf{k}\sigma'} - c_{b,-\textbf{k}\sigma}c_{a,\textbf{k}\sigma'}\bigr), \\
\hat{\Delta}^{s}_{a/b,\textbf{k}} &= \frac{1}{4}\sum_{\sigma\sigma'}[i\sigma_{2}]_{\sigma\sigma'}\bigl(c_{a,-\textbf{k}\sigma}c_{b,\textbf{k}\sigma'} + c_{b,-\textbf{k}\sigma}c_{a,\textbf{k}\sigma'}\bigr), \\ \hat{\Delta}^{s}_{a,\textbf{k}} &= \frac{1}{4}\sum_{\sigma\sigma'}[i\sigma_{2}]_{\sigma\sigma'}c_{a,-\textbf{k}\sigma}c_{a,\textbf{k}\sigma'}.
\end{aligned}
\end{equation}
where $\sigma_{i}$, $i=(1,2,3)$, are the Pauli matrices in spin space, $\hat{\textbf{d}}_{a/b,\textbf{k}}$ represents the interorbital spin-triplet pairing operators, and $\hat{\Delta}^{s}_{a/b,\textbf{k}}$($\hat{\Delta}^{s}_{a,\textbf{k}}$) represent the spin-singlet interorbital (intraorbital) pairings. As noted before \cite{klejnberg1999hund,SpalekPRB2001,HanPRB2004,Puetter2012EPL,vafek2017hund}, there is an attractive interorbital spin-triplet channel when $J_{H} > U'$, which will be our focus. Beyond the MF approach we use here, this form of interorbital spin-triplet pairing state has also been discovered in dynamical MF theory studies for Sr$_{2}$RuO$_{4}$, without the strict requirement that $J_{H} > U'$  \cite{Hoshino2015PRL,Hoshino2016PRB, Gingras2019PRL}.

Considering the specific case of $t_{2g}$ orbitals on a tetragonal lattice, these spin-triplet order parameters can be classified into different pairing channels according to the irreducible representations (irreps) of the $D_{4h}$ point group \cite{Ramires2019PRB, Suh2019}. The one-dimensional (1D) channels are defined as
\begin{equation}
\label{pairing_irreps}
    \begin{aligned}
        &\Delta_{A_{1g},1} = d^{x}_{xz/xy}+d^{y}_{xy/yz},\\[6pt]
        &\Delta_{A_{1g},2} = d^{z}_{yz/xz},\\[6pt]
        &\Delta_{B_{1g}} = d^{y}_{xy/yz}-d^{x}_{xz/xy},\\[6pt]
        &\Delta_{B_{2g}} = d^{y}_{xz/xy}+d^{x}_{xy/yz},\\[6pt]
        &\Delta_{A_{2g}} = d^{x}_{xy/yz}-d^{y}_{xz/xy},\\[6pt]
    \end{aligned}
\end{equation}
where $d^{i}_{a/b} = (U'-J_{H})\frac{1}{2N}\sum_{\textbf{k}}\langle{\hat{d}^{i}_{a/b,\textbf{k}}}\rangle$, $i=(x,y,z)$, and the two channels corresponding to a pairing state in the two-dimensional (2D) $E_{g}$ representation are $\displaystyle \{d_{yz/xz}^{x}, d_{yz/xz}^{y}\}$ and $\displaystyle \{d_{xz/xy}^{z}, d_{xy/yz}^{z}\}$ \cite{Ramires2019PRB,Suh2019}.

\section{\label{Three}interorbital Pairing In the Band Basis}

In a system with degenerate orbitals and Hund's coupling, the pairing occurs as a purely $s$-wave interband spin-triplet pairing in the band basis. However, in the more general case where the orbitals have different dispersions ($\xi_{\textbf{k}}^{a}\neq\xi_{\textbf{k}}^{b}$) and are coupled through interorbital hopping ($t_{\textbf{k}}^{a/b}$) and SOC, this is no longer the case. To understand the forms of pairing that can arise in such materials, we first review the importance of the SOC, which is seen clearly in a two-orbital model. Subsequently, we consider the case of three orbitals, focusing on the $t_{2g}$ case.

\subsection{Two orbitals}

We first review a two-orbital model with inversion and time-reversal symmetries. Defining $\psi_{\textbf{k}}^{\dagger} = (c^{\dagger}_{a,\textbf{k}\uparrow}, c^{\dagger}_{b,\textbf{k}\uparrow}, c^{\dagger}_{a,\textbf{k}\downarrow}, c^{\dagger}_{b,\textbf{k}\downarrow})$, which consists of creation operators for an electron in one of the two orbitals $a,b$ with spin $\sigma=\uparrow,\downarrow$, we use the Nambu spinor $\Psi_{\textbf{k}}^{\dagger} = (\psi_{\textbf{k}}^{\dagger}, \mathcal{T}\psi_{\textbf{k}}^{T}\mathcal{T}^{-1})$, where $\mathcal{T}$ refers to the time-reversal operator. Using the Pauli matrices and identity matrix, $\rho_{i},\sigma_{i}$, and $\tau_{i}~(i = 0,...3)$, represent the particle-hole, spin, and orbital bases, respectively. The Hamiltonian appears in this basis as
\begin{equation}
\begin{aligned}
    &H =\sum_{\textbf{k}} \Psi_{\textbf{k}}^{\dagger}(H_{0}(\textbf{k})+H_{\text{SOC}}(\textbf{k})+H_{\text{pair}})\Psi_{\textbf{k}}, \\
    &H_{0}(\textbf{k}) = \rho_{3}\sigma_{0}\left(\frac{\xi_{\textbf{k}}^{+}}{2}\tau_{0} + \frac{\xi_{\textbf{k}}^{-}}{2}\tau_{3} + t_{\textbf{k}}\tau_{1}\right),\\[6pt]
    &H_{\text{SOC}}(\textbf{k})=-\lambda_{\textbf{k}}\rho_{3}\sigma_{3}\tau_{2},\\[6pt]
    &H_\text{pair} = -d^{z} \rho_{2}\sigma_{3}\tau_{2}.
\end{aligned}
\end{equation}
The spin-independent part of the Hamiltonian, $H_{0}$, is generic for two orbitals, where $\xi_{\textbf{k}}^{\pm} = \xi_{\textbf{k}}^{a} \pm \xi_{\textbf{k}}^{b}$, and the interorbital hopping $t_{\textbf{k}}^{a/b}$ is denoted here by $t_{\textbf{k}}$. The SOC is chosen as $2\lambda_{\textbf{k}}L_{i}S_{z}$, where $L_{i}$ is the angular momentum matrix coupling orbitals $a$ and $b$, and $S_{z}$ represents the spin direction, chosen for simplicity. The allowed forms of SOC depend on the lattice symmetry, which we discuss below. The spin direction of the SOC pins the direction of the $d$ vector \cite{Puetter2012EPL, vafek2017hund}, resulting in the MF pairing, $d^{z}$ between orbitals $a$ and $b$, i.e., $d^{z}_{a/b}$.

Transforming the pairing to the band basis reveals the importance of the SOC, which can be done with the unitary transformation,
\begin{equation}    \label{first_trsf}
\begin{aligned}
\hspace{-2.5mm}
\begin{pmatrix}
   c_{a,\textbf{k}\sigma} \\[6pt]
   c_{b,\textbf{k}\sigma} \\
 \end{pmatrix}\hspace{-1.5mm}
 \setlength\arraycolsep{0.01mm}
 = \hspace{-1.5mm}\begin{pmatrix} 
    \frac{\eta_{\sigma}+1}{2}\tilde{f_{\textbf{k}}}-\frac{\eta_{\sigma}-1}{2}\tilde{f_{\textbf{k}}^{*}} & -\tilde{g_{\textbf{k}}}  \\[6pt]
  \tilde{g_{\textbf{k}}} & \frac{\eta_{\sigma}+1}{2}\tilde{f_{\textbf{k}}^{*}}-\frac{\eta_{\sigma}-1}{2}\tilde{f_{\textbf{k}}}
\end{pmatrix}\hspace{-2.5mm}
\begin{pmatrix}
   c_{\widetilde{\beta},\textbf{k}s} \\[6pt]
   c_{\widetilde{\alpha},\textbf{k}s} \\
 \end{pmatrix}\hspace{-1mm},
\end{aligned}
\end{equation}
where $\widetilde{\alpha},\widetilde{\beta}$ denote the bands, $s$ is a pseudospin index, $\eta_{\sigma}=\pm1$ for $\sigma=\uparrow,\downarrow$, and the coefficients of the transformation are $\tilde{f_{\textbf{k}}} = -\frac{(t_{\textbf{k}}+i\lambda_{\textbf{k}})}{\sqrt{t_{\textbf{k}}^2 + \lambda_{\textbf{k}}^2}}\sqrt{\frac{1}{2}(1 + \frac{\xi_{\textbf{k}}^{-}}{E_{1d,\textbf{k}}})}$ and $\tilde{g_{\textbf{k}}} = -\sqrt{\frac{1}{2}(1-\frac{\xi_{\textbf{k}}^{-}}{E_{1d,\textbf{k}}})}$, with $E_{1d,\textbf{k}}=\sqrt{(\xi_{\textbf{k}}^{-})^2 + 4(t_{\textbf{k}}^2+\lambda_{\textbf{k}}^2)}$. The pairing expressed in the band basis contains interband pseudospin-triplet and -singlet pairing. Our focus is on the intraband pseudospin-singlet pairing, as it is the dominant pairing. Thus we have
\begin{equation}    \label{pairing_two_orb}
\begin{aligned}
    \widetilde{H}_{\text{pair}} &= i\Delta_{\textbf{k}}\bigl[(c_{\widetilde{\beta},\textbf{k}+}c_{\widetilde{\beta},-\textbf{k}-} -c_{\widetilde{\beta},\textbf{k}-}c_{\widetilde{\beta},-\textbf{k}+}) \\[6pt]&- (c_{\widetilde{\alpha},\textbf{k}+}c_{\widetilde{\alpha},-\textbf{k}-} -c_{\widetilde{\alpha},\textbf{k}-}c_{\widetilde{\alpha},-\textbf{k}+})\bigr] + \text{H.c.},
\end{aligned}
\end{equation}
where
\begin{equation}
\Delta_{\textbf{k}} =  -2\frac{d^{z}\;\lambda_{\textbf{k}}}{E_{1d,\textbf{k}}},
\end{equation}
which is generated from the interorbital spin-triplet pairing only with nonzero SOC, $\lambda_{\textbf{k}}$. Additionally, the sign dependence throughout \textbf{k} space of the intraband pairing within the two-orbital model is purely given by the SOC, $\lambda_{\textbf{k}}$.

Note that the form of this result is not unique to the example of SOC with spin along the $z$ direction. For example, a SOC, $2L_{i}(\lambda_{\textbf{k}}^{x}S_{x} + \lambda_{\textbf{k}}^{y}S_{y})$, yields the same form for the intraband pairing $\sim (\lambda_{\textbf{k}}^{x}d^{x} + \lambda_{\textbf{k}}^{y}d^{y})$. Therefore, as pointed out previously \cite{Cheung2019PRB,Suh2019,clepkens2021}, starting from local interorbital pairing, \textbf{k}-SOC is a mechanism to obtain nontrivial gap structures. However, the possibilities depend on the form of SOC allowed by the symmetry. For example, in the case of a tetragonal lattice with $(a,b)=(d_{yz},d_{xz})$, there is the atomic SOC, $\lambda_{0}L_{z}S_{z}$, or other forms of nonlocal SOC with extended $s$-wave \textbf{k} dependence. However, there is also the possibility of SOC with a $d$-wave form factor, $\{\lambda_{\textbf{k}}^{x}L_{z}S_{x}, \lambda_{\textbf{k}}^{y}L_{z}S_{y}\}$, where $\lambda_{\textbf{k}}^{x}\sim \sin{k_{x}}\sin{k_{z}}$ and $\lambda_{\textbf{k}}^{y}\sim \sin{k_{y}}\sin{k_{z}}$, which results in a multicomponent $d$-wave pairing state with $\{d_{xz},d_{yz}\}$ components \cite{Cheung2019PRB}. At the level of two orbitals, the gap structure is determined by $\lambda_{\textbf{k}}$. This is no longer the case when there are three orbitals, as we show below.

\subsection{Three orbitals}

\begin{figure}[t!]
\includegraphics[width=86.5mm]{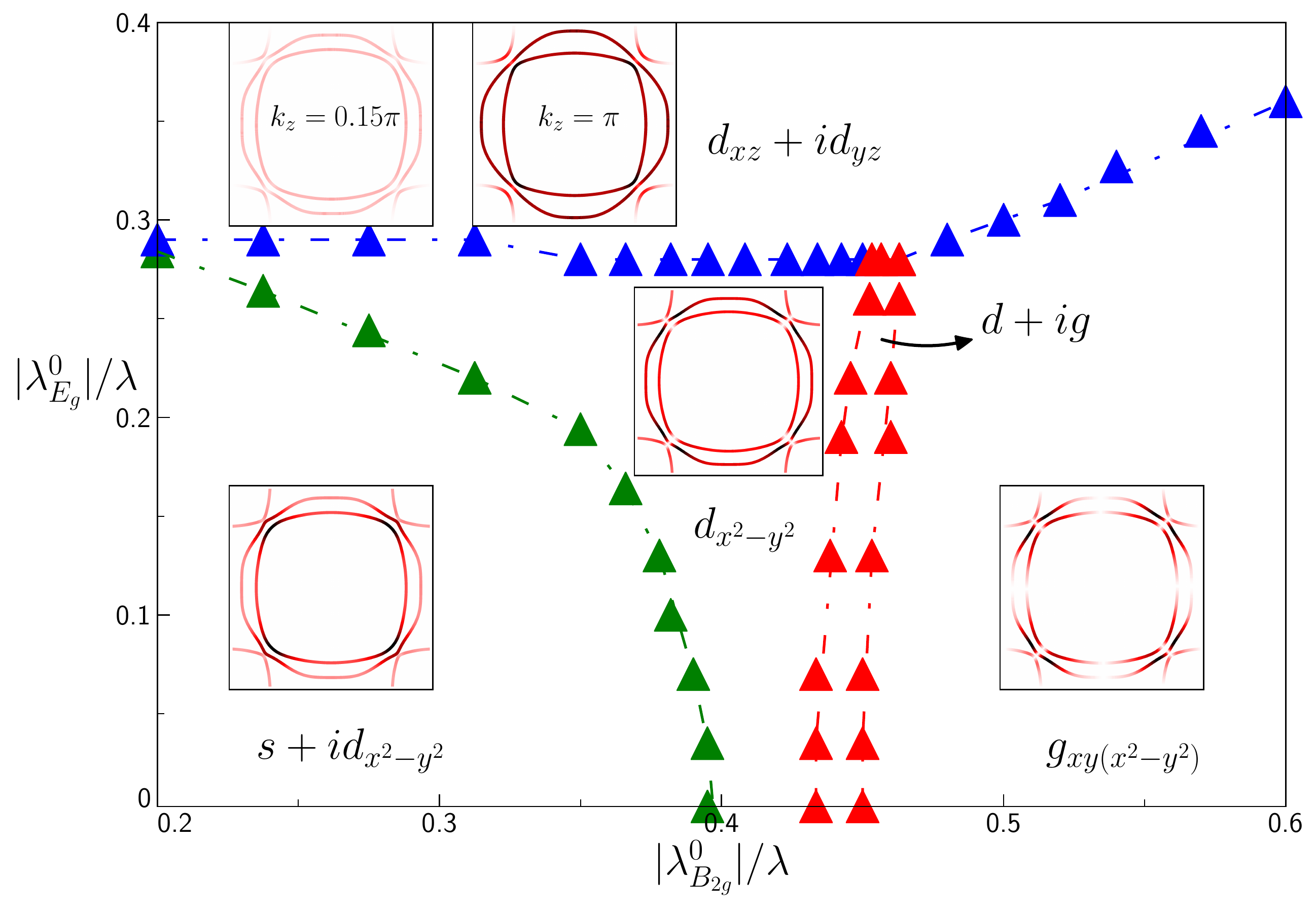}%
\caption{\label{fig1} Phase diagram at $T=0$ obtained from self-consistent MF theory by varying the SOC parameters, $\lambda$, $\lambda_{B_{2g}}^{0}$, and $\lambda_{E_{g}}^{0}$ within the full three-band model. The TB terms used are described in Sec.~\hyperref[Four]{IV}, along with the TB parameters in Table~\ref{TBparams_QE}, and the interaction Hamiltonian consists of the spin-triplet channels listed in Eqs.~(\ref{pairing_irreps}) with $\frac{J_{H}-U'}{2t_{1}}=0.74$. Along the $x$ axis, $|\lambda_{B_{2g}}^{0}|/\lambda$ is increased by decreasing $\lambda$ linearly while increasing $\lambda_{B_{2g}}^{0}$, such that $x=0.02$ corresponds to the DFT values with $\lambda=68.7$ meV, and at $x=$ 0.6, $\lambda=41.7$ meV. Along the $y$ axis, $|\lambda_{E_{g}}^{0}|$ is increased starting from the DFT value, $|\lambda_{E_{g}}^{0}|/\lambda=0.003$. The phase boundary continues as shown for $x<0.2$, with a slight increase in the critical value at which the $d_{xz}+id_{yz}$ state is stabilized. The inset in each phase shows the gap over the FS at each representative point, with the intensity indicating the size of the gap. For all states except $d_{xz}+id_{yz}$, the gap is shown at $k_{z}=0$ and normalized to the maximum at that $k_{z}$. For the $d_{xz}+id_{yz}$ state, the gap vanishes at $k_{z}=0$, and the gap is shown at both $k_{z}=0.15\pi$ and $k_{z}=\pi$, normalized to the maximum value at $k_{z}=\pi$.}
\end{figure}

Here we consider the three $t_{2g}$ orbitals in $D_{4h}$ symmetry. The SOC possibilities can be broken down into contributions based on the irrep of the operator part \cite{Ramires2019PRB}, $H_{\text{SOC}} = H_{\text{SOC}}^{A_{1g}} + H_{\text{SOC}}^{B_{1g}} + H_{\text{SOC}}^{B_{2g}} + H_{\text{SOC}}^{A_{2g}} + H_{\text{SOC}}^{E_{g}}$, where the form factor of the SOC must transform in the same way to respect the point-group symmetries. The atomic SOC is, $\displaystyle H_{\text{SOC}}^{A_{1g}} = i\lambda\sum_{\textbf{k},abc,\sigma\sigma'}\varepsilon_{abc}c_{a,\textbf{k}\sigma}^{\dagger}c_{b,\textbf{k}\sigma'}\sigma^{c}_{\sigma\sigma'}$, where $\varepsilon_{abc}$ is the completely antisymmetric tensor with $(a,b,c)=(yz,xz,xy)$ representing the $t_{2g}$ orbitals. The SOC in the $B_{2g}$ and $B_{1g}$ channels are
\begin{equation} \label{b2gSOC}
\begin{aligned}
    H_{\text{SOC}}^{B_{2g}} &= i\sum_{\textbf{k}\sigma\sigma'}\lambda_{\textbf{k}}^{B_{2g}} \sigma^{y}_{\sigma\sigma'}c^{\dagger}_{xz,\textbf{k}\sigma}c_{xy,\textbf{k}\sigma'} \\&-i\sum_{\textbf{k}\sigma\sigma'}\lambda_{\textbf{k}}^{B_{2g}} \sigma^{x}_{\sigma\sigma'}c^{\dagger}_{yz,\textbf{k}\sigma}c_{xy,\textbf{k}\sigma'} + \text{H.c.},
\end{aligned}
\end{equation}
and
\begin{equation} \label{b1gSOC}
\begin{aligned}
H_{\text{SOC}}^{B_{1g}} &= -i\sum_{\textbf{k}\sigma\sigma'}\lambda_{\textbf{k}}^{B_{1g}}\sigma^{y}_{\sigma\sigma'}c^{\dagger}_{yz,\textbf{k}\sigma}c_{xy,\textbf{k}\sigma'} \\ &-i\sum_{\textbf{k}\sigma\sigma'}\lambda_{\textbf{k}}^{B_{1g}}\sigma^{x}_{\sigma\sigma'}c^{\dagger}_{xz,\textbf{k}\sigma}c_{xy,\textbf{k}\sigma'} + \text{H.c.},
\end{aligned}
\end{equation}
where the form factors are, $\displaystyle \lambda_{\textbf{k}}^{B_{2g}}=4\lambda_{B_{2g}}^{0}\sin{k_x}\sin{k_y}$, and $\displaystyle \lambda_{\textbf{k}}^{B_{1g}} = 2\lambda_{B_{1g}}^{0}(\cos{k_x}-\cos{k_y})$. The forms of the SOC in the $A_{2g}$ and $E_{g}$ channels are listed in the Appendix. The contributions to the kinetic Hamiltonian from the orbital dispersions, $\xi_{\textbf{k}}^{a}$, and interorbital hoppings, $t_{\textbf{k}}^{a/b}$, are as defined in Eq.~(\ref{TBgeneral}). In a 3D model, $t_{\textbf{k}}^{a/b}$ is nonzero between all three $t_{2g}$ orbitals, and the form of these, along with the orbital dispersions are given in Sec.~\hyperref[Four]{IV}. The phase diagram of the full three-band numerical MF results for the pairing in terms of the spin-triplet channels defined in Eqs.~(\ref{pairing_irreps}) with the TB model presented in Sec.~\hyperref[Four]{IV} is shown in Fig.~\ref{fig1}. Here we tune $\lambda$, $\lambda_{B_{2g}}^{0}$, and $\lambda_{E_{g}}^{0}$ to show the competition among distinct pairing states. The details of this calculation are also given in Sec.~\hyperref[Four]{IV}.

\begin{figure*}[t!]
\includegraphics[width=120mm]{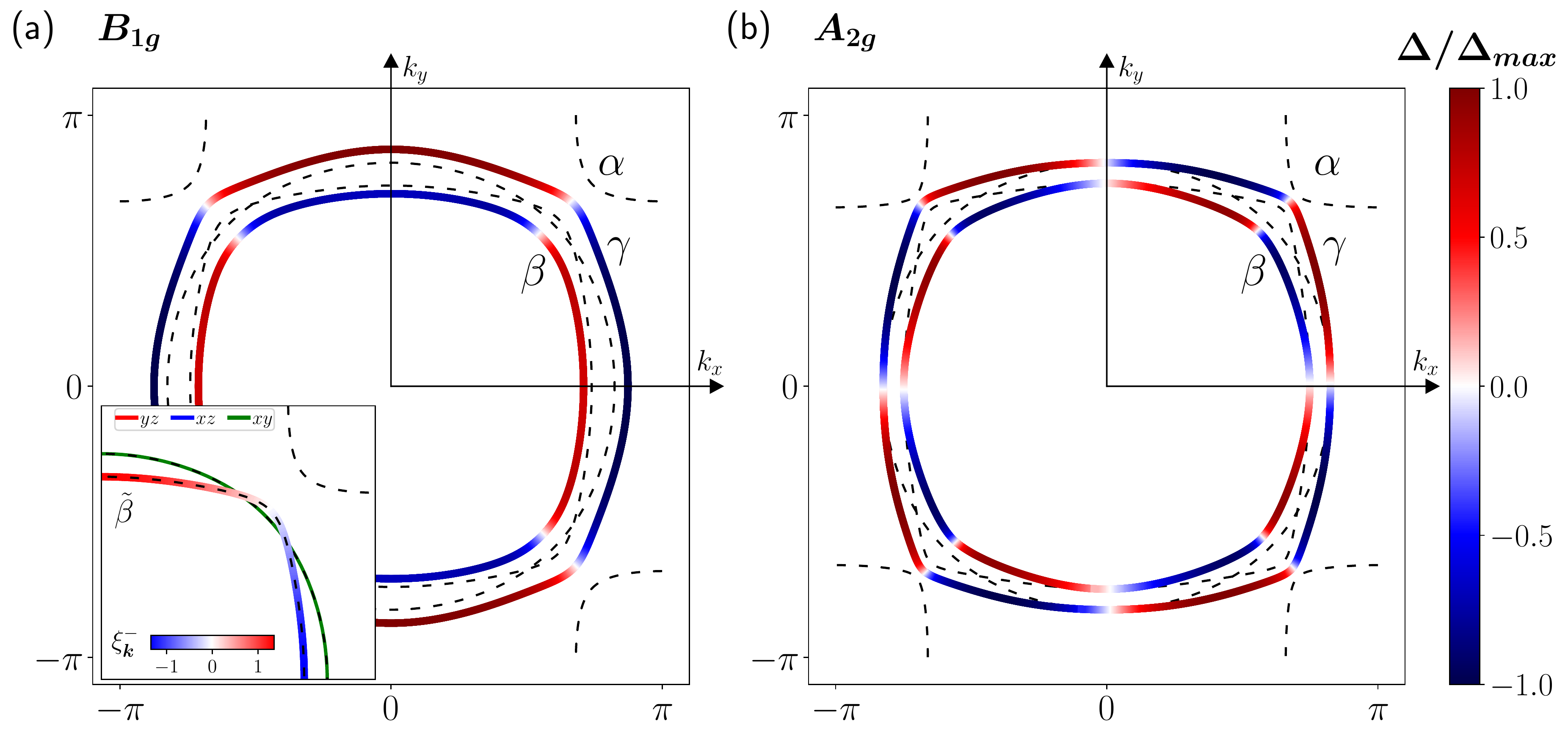}%
\caption{\label{fig2}Illustration of the gap in the two-band model, obtained using the pairing expressions in Eqs.~(\ref{expectation_atomic}) and (\ref{expectation_b2g}), with the dashed lines showing the unhybridized bands. (a) $B_{1g}$ pairing state with only the atomic SOC nonzero, $\lambda = 0.15$, $\lambda_{B_{2g}}^{0}=0$, and $\lambda_{B_{1g}}^{0}=0$. The gap vanishes along the diagonal directions reflecting the $d_{x^2-y^2}$ \textbf{k} dependence. The inset shows the unhybridized $\widetilde{\beta}$ and $d_{xy}$ bands, with the $d_{yz}/d_{xz}$ orbital content indicated by plotting $\xi_{\textbf{k}}^{-}$ over the $\widetilde{\beta}$ band. (b) $A_{2g}$ pairing state with only SOC in the $B_{2g}$ channel nonzero, $\lambda=0$, $\lambda_{B_{1g}}^{0}=0$, and $\lambda_{B_{2g}}^{0} = -0.05$. The gap vanishes along both the parallel and diagonal directions reflecting the $g_{xy(x^2-y^2)}$ \textbf{k} dependence. As shown in Eq.~(\ref{pair_brief}), the pairing occurs with a relative sign between the two bands in both pairing states.}
\end{figure*}

To offer insight into how higher angular momentum pairings are generated, we present an analytical analysis below. We approximate an effective two-band model by considering the $d_{yz}$ and $d_{xz}$ orbital mixing first, before incorporating the $d_{xy}$ orbital, and we neglect the 3D $E_{g}$ SOC and 3D hoppings. We define the basis $\psi_{\textbf{k}+(-)}^{\dagger}=(c_{yz,\textbf{k}\uparrow(\downarrow)}^{\dagger},c_{xz,\textbf{k}\uparrow(\downarrow)}^{\dagger},c_{xy,\textbf{k}\downarrow(\uparrow)}^{\dagger})$, in which the combined kinetic and SOC part of the Hamiltonian is now
\begin{equation}
\label{3orbMatrix}
\begin{aligned}
&H_{0}+H_{\text{SOC}}=\sum_{\textbf{k},s=\pm}\psi_{\textbf{k}s}^{\dagger}A_{\textbf{k}s}\psi_{\textbf{k}s},\\[6pt]
&A_{\textbf{k}s} = 
 \begin{pmatrix} 
    \xi_{\textbf{k}}^{yz} & t_{\textbf{k}}+is\lambda & \lambda_{Y,s}(\textbf{k}) \\[7pt]
    t_{\textbf{k}}-is\lambda & \xi_{\textbf{k}}^{xz} &\lambda_{X,s}(\textbf{k}) \\[7pt]
    \lambda_{Y,s}^{*}(\textbf{k}) & \lambda_{X,s}^{*}(\textbf{k}) & \xi_{\textbf{k}}^{xy}
\end{pmatrix},
\end{aligned}
\end{equation}
where $\displaystyle \lambda_{Y,s}(\textbf{k}) = -s\lambda-s\lambda_{\textbf{k}}^{B_{1g}}-i\lambda_{\textbf{k}}^{B_{2g}}$ and $\displaystyle \lambda_{X,s}(\textbf{k}) = i\lambda -i\lambda_{\textbf{k}}^{B_{1g}} + s\lambda_{\textbf{k}}^{B_{2g}}$. The interorbital hopping is $t_{\textbf{k}} = t_{\textbf{k}}^{yz/xz
}=-4t_{1d}\sin{k_x}\sin{k_y}$, and the orbital dispersions are generically chosen as $\xi_{\textbf{k}}^{yz/xz} = -2t_{1}\cos{k_{y/x}}-2t_{2}\cos{k_{x/y}}-\mu_{1d}$, $\xi_{\textbf{k}}^{xy} = -2t_{3}(\cos{k_x}+\cos{k_y})-4t_{4}\cos{k_x}\cos{k_y}-\mu_{xy}$. As noted, we first consider the $(d_{yz},d_{xz})$ band mixing and then incorporate the $d_{xy}$ orbital through SOC. We thus transform the Hamiltonian to a model consisting of one of the bands from the mixing between $(d_{yz},d_{xz})$ through SOC and $t_{\textbf{k}}$, and the remaining $d_{xy}$ orbital. These intermediate bands from the mixing in the $(d_{yz}$, $d_{xz})$ subspace are denoted by $(\widetilde{\alpha}, \widetilde{\beta})$, to distinguish them from the $(\alpha, \beta)$ bands obtained after including the coupling to the $d_{xy}$ orbital. The result is the band dispersions, $\xi_{\textbf{k}}^{\widetilde{\alpha}}$ and $\xi_{\textbf{k}}^{\widetilde{\beta}}$, as well as new effective SOC terms between these bands and the $d_{xy}$ orbital.  Therefore, we apply Eq.~(\ref{first_trsf}), with $a=d_{yz}$,$~b=d_{xz}$, to the above Hamiltonian, and project out the $\widetilde{\alpha}$ band. In this basis, with $\widetilde{\psi}_{\textbf{k}+(-)}^{\dagger}=(c_{\widetilde{\beta},\textbf{k}\uparrow(\downarrow)}^{\dagger}, c_{xy,\textbf{k}\downarrow(\uparrow)}^{\dagger})$, the kinetic Hamiltonian is
\begin{equation}
\begin{aligned}
&\widetilde{H}_{0}=\sum_{\textbf{k},s=\pm}\widetilde{\psi}_{\textbf{k}s}^{\dagger}\widetilde{A}_{\textbf{k}s}\widetilde{\psi}_{\textbf{k}s},\\[6pt]
&\widetilde{A}_{\textbf{k}s} = 
 \begin{pmatrix} 
    \xi_{\textbf{k}}^{\widetilde{\beta}} & s\eta_{\widetilde{\beta},\textbf{k}}^{R}+i\eta_{\widetilde{\beta},\textbf{k}}^{I}  \\[7pt]
     s\eta_{\widetilde{\beta},\textbf{k}}^{R}-i\eta_{\widetilde{\beta},\textbf{k}}^{I} & \xi_{\textbf{k}}^{xy}
\end{pmatrix},
\end{aligned}
\end{equation}
where the real, $\eta_{\widetilde{\beta},\textbf{k}}^{R}$, and imaginary, $\eta_{\widetilde{\beta},\textbf{k}}^{I}$, parts of the effective SOC between the $\widetilde{\beta}$ band and the $d_{xy}$ orbital are
\begin{equation}
\label{effectiveSOC}
\begin{aligned}
\eta_{\widetilde{\beta},\textbf{k}}^{I} &= -|\tilde{g_{\textbf{k}}}|(\lambda - \lambda^{B_{1g}}_{\textbf{k}}) \hspace{-0.3mm}  \\[6pt]&\hspace{-2mm}+ \frac{|\tilde{f_{\textbf{k}}}|}{\sqrt{t_{\textbf{k}}^{2}+\lambda^2}}(-\lambda^2 \hspace{-0.6mm} - \hspace{-1.0mm}\lambda \lambda^{B_{1g}}_{\textbf{k}}\hspace{-0.6mm} +\hspace{-0.8mm} \lambda^{B_{2g}}_{\textbf{k}}t_{\textbf{k}}), \\[6pt]
\eta_{\widetilde{\beta},\textbf{k}}^{R} &= -|\tilde{g_{\textbf{k}}}|\lambda^{B_{2g}}_{\textbf{k}} \\[6pt]&\hspace{-2mm}+ \frac{|\tilde{f_{\textbf{k}}}|}{\sqrt{t_{\textbf{k}}^2+\lambda^2}}(\lambda t_{\textbf{k}} + \lambda \lambda^{B_{2g}}_{\textbf{k}} + \lambda^{B_{1g}}_{\textbf{k}}t_{\textbf{k}} ).
\end{aligned}
\end{equation}
The imaginary part has both an anisotropic $s$-wave \textbf{k} dependence, but also a $d_{x^2-y^2}$ dependence due to $\lambda_{\textbf{k}}^{B_{1g}}$. Interestingly, the real part has both a $d_{xy}$ component given directly by $\lambda_{\textbf{k}}^{B_{2g}}$, and indirectly through terms like $\lambda\;t_{\textbf{k}}$, but also a $g$-wave dependence appearing through $\lambda_{\textbf{k}}^{B_{1g}}t_{\textbf{k}}\sim \sin{k_x}\sin{k_y}(\cos{k_x}-\cos{k_y})$. 

The effective SOC now mixes the $\widetilde{\beta}$ band with the $d_{xy}$ orbital, so another transformation is required to diagonalize the above Hamiltonian, which is accomplished by
\begin{equation} \label{last_trsf}
\begin{aligned}
\begin{pmatrix}
   c_{\widetilde{\beta},\textbf{k}\sigma} \\[6pt]
   c^{xy}_{\textbf{k}-\sigma} \\
 \end{pmatrix}\hspace{-1.5mm}
 \setlength\arraycolsep{0.005mm}
 = \hspace{-1.5mm}\begin{pmatrix} 
    \frac{\eta_{\sigma}+1}{2}f_{\textbf{k}}+\frac{\eta_{\sigma}-1}{2}f_{\textbf{k}}^{*} & \eta_{\sigma}g_{\textbf{k}}  \\[6pt]
  g_{\textbf{k}} & -\frac{\eta_{\sigma}+1}{2}f_{\textbf{k}}^{*}+\frac{\eta_{\sigma}-1}{2}f_{\textbf{k}}
\end{pmatrix}\hspace{-2.5mm}
\begin{pmatrix}
   c_{\beta,\textbf{k}s} \\[6pt]
   c_{\gamma,\textbf{k}s} \\
 \end{pmatrix},
\end{aligned}
\end{equation}
with $s=+(-)$ for $\sigma=\uparrow(\downarrow)$ and now $f_{\textbf{k}} = -\frac{\eta_{\widetilde{\beta},\textbf{k}}}{|\eta_{\widetilde{\beta},\textbf{k}}|}\sqrt{\frac{1}{2}(1 + \frac{\widetilde{\xi}_{\textbf{k}}^{-}}{E_{\widetilde{\beta},\textbf{k}}})}$, $g_{\textbf{k}} = -\sqrt{\frac{1}{2}(1-\frac{\widetilde{\xi}_{\textbf{k}}^{-}}{E_{\widetilde{\beta},\textbf{k}}})}$, and we have defined $\widetilde{\xi}_{\textbf{k}}^{-} = \xi_{\textbf{k}}^{\widetilde{\beta}}-\xi_{\textbf{k}}^{xy}$, as well as $E_{\widetilde{\beta},\textbf{k}} = \sqrt{(\widetilde{\xi}_{\textbf{k}}^{-})^2 + 4|\eta_{\widetilde{\beta},\textbf{k}}|^2}$. With this, the intraband pairing operator in terms of spin-triplet pairings between the original $\widetilde{\beta}$ band and the $d_{xy}$ orbital is
\begin{equation}    \label{intra_general}
\begin{aligned}
    &\langle{c_{\beta,-\textbf{k}+}c_{\beta,\textbf{k}-} - c_{\beta,-\textbf{k}-}c_{\beta,\textbf{k}+}}\rangle = \\[6pt] &\frac{-4i\bigl[\eta_{\widetilde{\beta},\textbf{k}}^{R}\langle{\hat{d}_{\textbf{k}}^{y}}\rangle + \eta_{\widetilde{\beta},\textbf{k}}^{I}\langle{\hat{d}_{\textbf{k}}^{x}}\rangle \bigr]}{E_{\widetilde{\beta},\textbf{k}}} + ...~,
\end{aligned}
\end{equation}
where $...$ indicates contributions from pairing between $\widetilde{\beta}$ and $\widetilde{\beta}$, as well as between $d_{xy}$ and $d_{xy}$, not considered here. Note that the result is identical to that found in the two orbital model: the intraband pairing is proportional to the SOC, except that here it is the effective SOC, $\eta_{\widetilde{\beta},\textbf{k}}$, that appears, which contains \textbf{k} dependence given by products of the various SOC and TB terms, as shown in Eq.~(\ref{effectiveSOC}). However, the pairing operators appearing in this expression are themselves \textbf{k}-dependent mixtures of the pairing between $d_{yz}/d_{xz}$ and $d_{xy}$ orbitals, given by Eq.~(\ref{first_trsf}).

To highlight the effect of three orbitals, we consider two limiting cases for the intraband pairing, which in both of these appear as
\begin{equation}    \label{pair_brief}
\begin{aligned}
    \widetilde{H}_{\text{pair}} = i&\Delta_{\textbf{k}}\bigl[(c_{\beta,\textbf{k}+}c_{\beta,-\textbf{k}-} -c_{\beta,\textbf{k}-}c_{\beta,-\textbf{k}+}) \\[6pt]&- (c_{\gamma,\textbf{k}+}c_{\gamma,-\textbf{k}-} -c_{\gamma,\textbf{k}-}c_{\gamma,-\textbf{k}+})\bigr] + \text{H.c.}.\\[6pt]
\end{aligned}
\end{equation}
The general result for the pairing including all relevant spin-triplet channels is given in the Appendix.  Including only the atomic SOC, $\lambda$, we find that $\Delta_{\textbf{k}}$ contains the contribution, 
\begin{equation}
\label{expectation_atomic}
\begin{aligned}
      \frac{1}{E_{\widetilde{\beta},\textbf{k}}}{\Delta_{B_{1g}}}{\frac{{\lambda\;\xi_{\textbf{k}}^{-}}}{E_{1d,\textbf{k}}}} \propto \lambda\;(\cos{k_x}-\cos{k_y}).
\end{aligned}
\end{equation}
This has a $d$-wave symmetry due to $\xi_{\textbf{k}}^{-} = \xi^{yz}_{\textbf{k}}-\xi^{xz}_{\textbf{k}}$.  It contains nodes along the $k_{x}=\pm k_{y}$ directions and occurs as an intraband pairing due to the variation of $d_{yz}/d_{xz}$ orbital character in the $\widetilde{\beta}$ band as a result of $\xi_{\textbf{k}}^{-}$, which is then coupled with the $d_{xy}$ orbital through $\lambda$. We emphasize that the origin of the $d_{x^2-y^2}$ character here is from the two components of the $d$ vector, $d_{xz/xy}^{x}$ and $d_{xy/yz}^{y}$ occurring with different signs, i.e., $d_{xz/xy}^{x}=-d_{xy/yz}^{y}$, resulting in a $B_{1g}$ pairing state with the $\frac{\pi}{2}$ rotational symmetry broken. The competition between the $B_{1g}$ ($d_{x^2-y^2}$) and $A_{1g}$ ($s$) pairing states depends on the details of the kinetic terms and is discussed in the Appendix. 


Alternatively, we can consider the case where only the SOC in the $B_{2g}$ channel, $\lambda_{\textbf{k}}^{B_{2g}}$, is nonzero, favoring the order parameters between the same orbitals as before but with the $d$-vector directions switched due to the form of the $B_{2g}$ SOC. Then the pairing has a contribution, 
\begin{equation}
\label{expectation_b2g}
\begin{aligned}
     &\frac{1}{E_{\widetilde{\beta},\textbf{k}}}{\Delta_{A_{2g}}}\frac{{\lambda_{\textbf{k}}^{B_{2g}}\xi_{\textbf{k}}^{-}}}{E_{1d,\textbf{k}}}\\[6pt]& \propto \lambda_{B_{2g}}^{0}\sin{k_x}\sin{k_y}(\cos{k_x}-\cos{k_y}).
\end{aligned}
\end{equation}
The result is $g$-wave \textbf{k} dependence of the form, $\sin{k_x}\sin{k_y}(\cos{k_x}-\cos{k_y})$, due to $\lambda_{\textbf{k}}^{B_{2g}}\xi_{\textbf{k}}^{-}$. The pairing therefore contains nodes along both the $k_{x}=\pm k_{y}$ and $k_{x(y)}=0$ directions. As in the previous case, the competition between the $A_{2g}$ ($g_{xy(x^2-y^2)}$) and $B_{2g}$ ($d_{xy}$) pairing states depends on the details of the kinetic terms (see Appendix).


To illustrate the gap in these pairing states, we plot the intraband pairing expressions given in  Eqs.~(\ref{expectation_atomic}) and (\ref{expectation_b2g}) in the two-band model over the FS generated with a set of TB parameters. The result is shown in Fig.~\ref{fig2}, for (a) a $d_{x^2-y^2}$ ($B_{1g}$) pairing state with $\lambda=0.15$, $\lambda_{B_{2g}}^{0}=0$, $\lambda_{B_{1g}}^{0}=0$ and (b) a $g_{xy(x^2-y^2)}$ ($A_{2g}$) pairing state with $\lambda=0$, $\lambda_{B_{2g}}^{0}=-0.05$, $\lambda_{B_{1g}}^{0}=0$. The inset in Fig.~\ref{fig2} (a) shows the orbital dispersion difference on the $\widetilde{\beta}$ band which gives rise to the $d_{x^2-y^2}$ symmetry when the atomic SOC is nonzero. When \textbf{k}-SOC is included, as in (b), there is an extra \textbf{k} dependence, such as $(\sin{k_x}\sin{k_y})$, which allows for the possibility of a $g$-wave gap in the $A_{2g}$ pairing state.

With three orbitals coupled through the atomic SOC, $\lambda$, and the SOCs with $B_{2g}$, $\lambda_{\textbf{k}}^{B_{2g}}$, and $B_{1g}$, $\lambda_{\textbf{k}}^{B_{1g}}$, form factors in Eq.~(\ref{3orbMatrix}), the intraband pairing contains contributions in each of the 1D channels in Eqs.~(\ref{pairing_irreps}). The full expression for the intraband pairing in this case, as well as for a similar analysis with the 3D $E_{g}$ SOC instead is given in the Appendix. The dominant pairing among these channels depends on the balance between the kinetic and SOC terms due to their effect on each channel's projection onto the intraband pairing. This is also seen through the commutation and anti-commutation relations with the pairing Hamiltonian \cite{fischer2013NJP, Ramires2016PRB, Ramires2018PRB}. Various nontrivial pairing states including the $d_{x^2-y^2}$ and $g_{xy(x^2-y^2)}$ states illustrated above can thus be close in energy, making a multicomponent order parameter composed of pairings with distinct symmetries plausible. This can be seen in the phase diagram for the full three-orbital model, shown in Fig.~\ref{fig1}, for which we now discuss the details.

\section{\label{Four}Application to \protect\Sr}

Considering the specific example of Sr$_{2}$RuO$_{4}$, we use a three-band microscopic model composed of the $t_{2g}$ orbitals, and numerically examine the effect of the SOC on the competing pairing states within self-consistent MF theory at zero temperature. We include the \textbf{k}-SOC terms and also further-neighbour and out-of-plane hoppings in the model for completeness. The orbital dispersions included in the general kinetic Hamiltonian, $H_{0}$, given by Eq.~(\ref{TBgeneral}) are
\begin{equation}
    \begin{aligned}
    \xi_{\textbf{k}}^{xz/yz} = &-2t_{1}\cos{k_{x/y}} - 2t_{2}\cos{k_{y/x}} - 2t_{1}^{'}\cos{2k_{x/y}} \\&-2t_{2}^{'}\cos{2k_{y/x}} -2t_{1}^{''}\cos{3k_{x/y}} \\&+ 4t_{4}\cos{k_x}\cos{k_y}-4t_{4}^{'}\cos{2k_{x/y}}\cos{k_{y/x}} \\&- 4t_{4}^{''}\cos{2k_{y/x}}\cos{k_{x/y}} \\&+ 8t_{z}^{1d}\cos{\frac{k_x}{2}}\cos{\frac{k_y}{2}}\cos{\frac{k_z}{2}} - \mu_{1d},
    \end{aligned}
\end{equation}
\begin{equation}
    \begin{aligned}
    \xi_{\textbf{k}}^{xy} = &-2t_{3}(\cos{k_x}+\cos{k_y}) -2t_{3}^{'}(\cos{2k_x}+\cos{2k_y})\\
    &-2t_{3}^{''}(\cos{3k_x}+\cos{3k_y}) -4t_{5}\cos{k_x}\cos{k_y}\\
    &-4t_{5}^{'}(\cos{2k_{x}}\cos{k_y}+\cos{2k_{y}}\cos{k_{x}})\\
    &+8t_{z}^{xy}\cos{\frac{k_x}{2}}\cos{\frac{k_y}{2}}\cos{\frac{k_z}{2}} - \mu_{xy},
    \end{aligned}
\end{equation}
and the interorbital hoppings are
\begin{equation}    \label{interorbhoppings}
\begin{aligned}
    &t_{\textbf{k}}^{yz/xz} = -4t_{6}\sin{k_x}\sin{k_y} -8t_{7}\sin{\frac{k_x}{2}}\sin{\frac{k_y}{2}}\cos{\frac{k_z}{2}}\\
    &\hspace{12mm}-4t_{6}^{'}(\sin{2k_x}\sin{k_y}+\sin{2k_y}\sin{k_x}),\\
    &t_{\textbf{k}}^{yz/xy} = -8t_{8}\sin{\frac{k_x}{2}}\cos{\frac{k_y}{2}}\sin{\frac{k_z}{2}},\\
    &t_{\textbf{k}}^{xz/xy} = -8t_{8}\cos{\frac{k_x}{2}}\sin{\frac{k_y}{2}}\sin{\frac{k_z}{2}}.
\end{aligned}
\end{equation}
The SOC part of the Hamiltonian is given by $\displaystyle H_{\text{SOC}} = H_{\text{SOC}}^{A_{1g}} + H_{\text{SOC}}^{B_{1g}} + H_{\text{SOC}}^{B_{2g}} + H_{\text{SOC}}^{E_{g}}$, where $H_{\text{SOC}}^{A_{1g}}$ denotes the atomic SOC as before. The SOC in the $B_{2g}$ and $B_{1g}$ channels are given in Eqs.~(\ref{b2gSOC}) and (\ref{b1gSOC}), and the form of the SOC in the $E_{g}$ channel is given in the Appendix. 


\begin{figure}[t!]
\includegraphics[width=60mm]{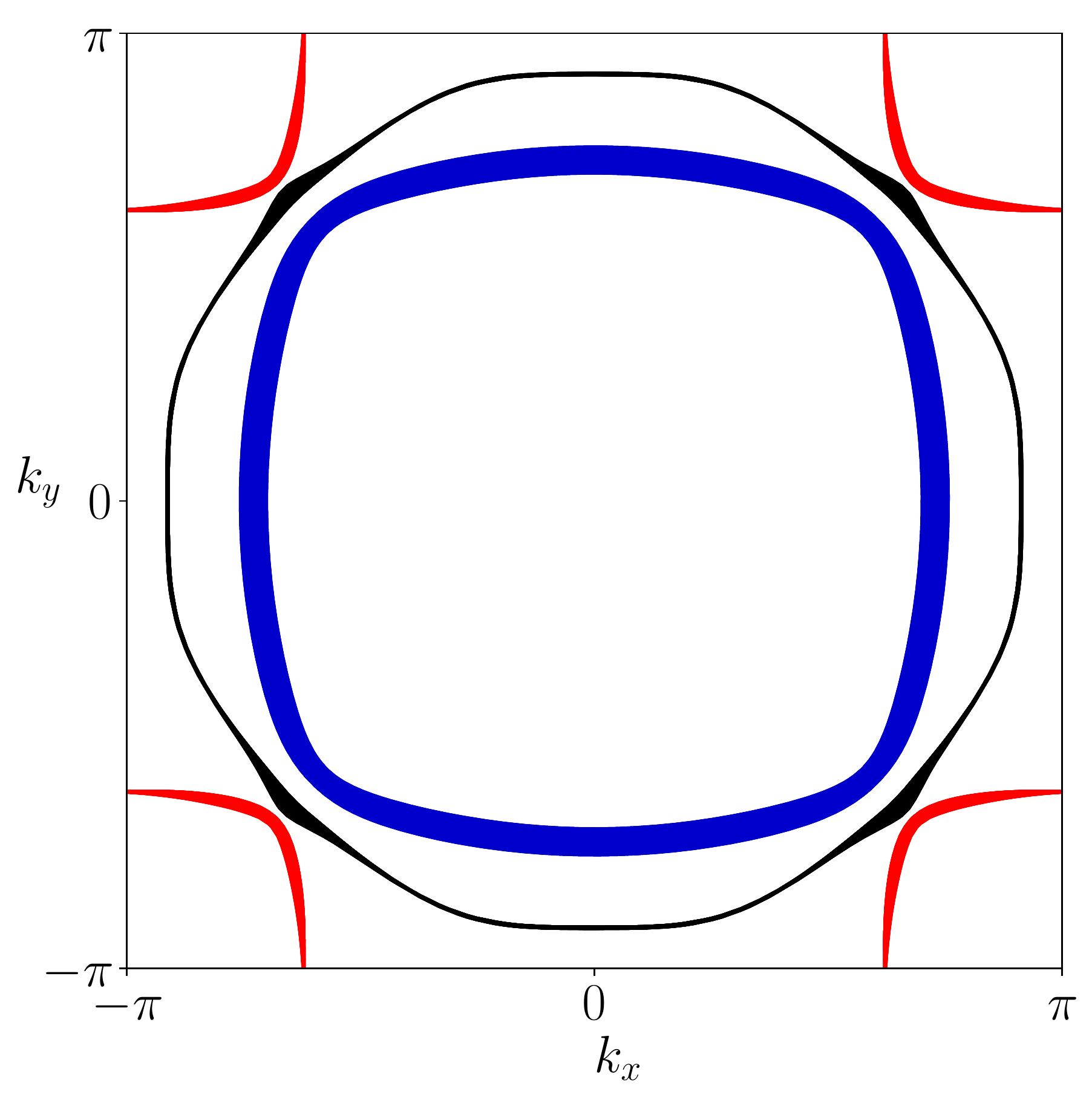}%
\caption{\label{fig3}FS obtained with the TB model presented in the main text, with TB parameters given in Table~\ref{TBparams_QE}. The contours at each $k_{z}$ in the Brillouin zone (BZ) are projected onto the $k_{x}-k_{y}$ plane, resulting in a varying thickness such that the thicker parts of the contours correspond to a larger $k_{z}$ dispersion.}
\end{figure}

To obtain the TB parameters, we perform density functional theory (DFT) calculations with the QUANTUM ESPRESSO (QE) package \cite{Giannozzi_2009, Giannozzi_2017} using the fully relativistic optimized norm-conserving Vanderbilt \cite{HamannPRB2017} pseudopotential. The energy cutoff of the plane-wave basis is 60 Ry. The relevant TB parameters are obtained using WANNIER90 \cite{MOSTOFI2014}. The parameter set based on the QE calculations is given in Table~\ref{TBparams_QE}, and the 3D FS calculated from the resulting TB model and projected onto the $k_{x}-k_{y}$ plane is shown in Fig.~\ref{fig3}. We also obtain a TB set in the same way based on the Vienna \textit{ab initio} simulation package (VASP) \cite{KressePRB1993}, using the projector augmented wave potential \cite{BlochlPRB1994} and the Perdew-Burke-Ernzerhof exchange-correlation functional \cite{PerdewPRL1996}, with energy cutoff of 400 eV. The parameters based on the VASP calculations, as well as those introduced in Ref.~\cite{Roising2019prr} with the \textbf{k}-SOC parameters of Table~\ref{TBparams_QE} are used as a comparison for the main results discussed below.

\begin{table}[h!]
\centering
\caption{TB parameters obtained through the DFT calculations. All parameters are in units of meV. The atomic SOC, $\lambda$, is the average of the SOC parameter between the $(d_{yz},d_{xz})$ orbitals and the $(d_{yz}/d_{xz},d_{xy})$ orbitals, which are slightly different due to the tetragonal anisotropy.}
 \begin{tabular}{| m{0.75cm}  m{0.75cm} m{0.75cm}  m{0.75cm}  m{0.75cm}  m{0.75cm}  m{0.75cm} m{0.75cm} m{0.75cm} m{0.75cm} |}
 \hline\hline
 $t_{1}$ & $t_{1}^{'}$ & $t_{1}^{''}$ & $t_{2}$ & $t_{2}^{'}$ & $t_{4}$ & $t_{4}^{'}$ & $t_{4}^{''}$ & $t_{z}^{1d}$ & $\mu_{1d}$ \\  
 \hline
 412.4 & 1.5 & 11.2 & 50.8 & 0.6 & 10.6 & -11.6 & 0.8 & -20.0 & 415.5  \\ 
 \hline
 $t_{3}$ & $t_{3}^{'}$ & $t_{3}^{''}$ & $t_{5}$ & $t_{5}^{'}$ & $t_{z}^{xy}$ &  $\mu_{xy}$ & $t_{6}$ & $t_{6}^{'}$ & $t_{7}$\\
 \hline
  402.5 & 1.3 & -0.2 & 142.7 & 21.6 & 1.8 & 511.1  & 12.4 & 2.5 & -8.7 \\ 
  \hline
  $t_{8}$ & $\lambda$ & $\lambda_{B_{2g}}^{0}$ & $\lambda_{B_{1g}}^{0}$ & $\lambda_{E_{g}}^{0}$ & $\lambda_{E_{g},1d}^{0}$ &  & & &\\
  \hline
  7.0 & 68.7 & -1.3 & -0.3 & -0.2 & 0.9 & & & & \\ 
 \hline\hline
\end{tabular}
\label{TBparams_QE}
\end{table}

The spin-triplet pairing terms given in Eq.~(\ref{interactions}) are included at the MF level, and the nine complex local spin-triplet order parameters in the orbital basis, given in Eq.~(\ref{OPs}), are calculated self-consistently, with $\frac{J_{H}-U'}{2t_{1}}=0.74$, focusing on the effects of the atomic SOC, $\lambda$, and the SOC in the $B_{2g}$, $\lambda_{B_{2g}}^{0}$, and $E_{g}$, $\lambda_{E_{g}}^{0}$, channels. We note that the spin-triplet order parameters correspond to both intraband and interband pairing in the band basis, as discussed previously. The pairing states are classified according to the channels given in Eqs.~(\ref{pairing_irreps}), however we label them as $s, d_{x^2-y^2}$, etc., which corresponds to the \textbf{k} dependence of the pairing in the band basis and represents the transformation properties of the corresponding irrep. Using the TB set in Table~\ref{TBparams_QE}, we find the favoured pairing to be a time-reversal symmetry breaking $A_{1g}+iB_{1g}$ ($s+id_{x^2-y^2})$ state, with a dominant $s$ component. Note that the pairing in both of these channels is made of the spin-triplet order parameters, $(d_{xz/xy}^{x}, d_{xy/yz}^{y})$, thus an $s+id_{x^2-y^2}$ state corresponds to $\text{Re}(d_{xz/xy}^{x})=\text{Re}(d_{xy/yz}^{y})$ and $\text{Im}(d_{xz/xy}^{x})=-\text{Im}(d_{xy/yz}^{y})$, up to the overall phase. 

Since SOC can be further enhanced by correlations \cite{LiuPRL2008,IsobePRB2014,zhang2016PRL,Kim2018PRL,Tamai2019PRX}, we study the effects of SOC on the pairing by tuning $\lambda$, $\lambda_{B_{2g}}^{0}$, and $\lambda_{E_{g}}^{0}$, starting from the initial DFT values. Increasing $\lambda_{B_{1g}}^{0}$ has a significantly smaller effect than $\lambda_{B_{2g}}^{0}$, and thus not shown here. At each point in the parameter space, the order parameters are initialized in different pairing channels and the energies are compared if multiple self-consistent solutions are found. The resulting phase diagram is shown in Fig.~\ref{fig1}, where $|\lambda_{B_{2g}}^{0}|/\lambda\equiv x$ is plotted along the $x$ axis, with $\lambda$ decreasing linearly with $x$ such that $x=0.02$ corresponds to the DFT value of $\lambda=68.7$ meV (far left of the phase diagram not shown in Fig.~\ref{fig1} for clarity of the figure; the pairing is $s+id_{x^2-y^2}$ as discussed above). At the right end of the $x$ axis, $\lambda=41.7$ meV, with $|\lambda_{B_{2g}}^{0}|$ increasing simultaneously. At a fixed value of $x$, $|\lambda_{E_{g}}^{0}|/\lambda\equiv y$ is increased along the $y$ axis by increasing $|\lambda_{E_{g}}^{0}|$, starting from the DFT value of $y=0.003$.

As shown in Fig.~\ref{fig1}, depending on the SOC there are several competing pairing states with distinct symmetry that can coexist at $T=0$.  If the $\textbf{k}$-SOC parameters are both small compared to $\lambda$, then we find the aforementioned $s+id_{x^2-y^2}$ pairing state.  The $d_{xy}$ ($B_{2g}$) pairing in $s+id_{xy}$ discussed in Ref.~\cite{clepkens2021} is found to be subleading due to the large atomic SOC, orbital hopping difference, and balance between the interorbital hopping terms. The TB set used here contains similar values of $t_{6}$ and $t_{7}$, compared to $t_{7}\gg t_{6}$ in Ref.~\cite{clepkens2021} which has a favourable effect on the $d_{xy}$ pairing, as seen by the pairing expression given in the Appendix. The competition between either $B_{1g}$ or $A_{2g}$ pairings, and the $B_{2g}$ pairing thus depends on the value of $(t_{1}-t_{2})$, $\lambda$, and the interorbital hopping parameters, however, they tend to be close in energy.

Near $x=0$, the $A_{1g}$ component is larger than the $B_{1g}$, but as $x$ is increased, the $B_{1g}$ component increases, with the $A_{1g}$ and $B_{1g}$ gap components being approximately the same size at $x\approx0.3$. If $x$ is increased past the point of degeneracy, the dominant pairing state eventually becomes a single component $B_{1g}$ state for $x\gtrsim0.39$, and increasing $y$ can also make the $B_{1g}$ component dominant, depending on the value of $x$. Interestingly, this state eventually gives way to an $A_{2g}$ ($g_{xy(x^2-y^2)}$) pairing state along the $x$ axis for $x\gtrsim0.45$, however, there is a small window of coexistence in which the dominant pairing state is, $B_{1g} + iA_{2g}$ ($d_{x^2-y^2}+ig_{xy(x^2-y^2})$, for $x\gtrsim0.43$. This pairing was one of the suggested pairings for Sr$_{2}$RuO$_{4}$ \cite{Kivelson2020npj, yuan2021}. At any value of $x$, for large enough $y$, there is a transition to the $E_{g}$ ($d_{xz}+id_{yz}$) pairing state found in Ref.~\cite{Suh2019} with the critical $y$ value ranging from $\approx 0.28$ to $0.36$.


To compare to the results presented above, we have also performed calculations as a function of $x$ with both the TB parameters based on the VASP DFT calculations, and the parameters in Ref.~\cite{Roising2019prr} found through a Monte Carlo fitting to the DFT results of Ref.~\cite{Veenstra2014PRL}. Both of these are similar, but with an important difference from the parameters listed above being a smaller value for $(t_{1}-t_{2})$, of $246$ and $230$ meV in the VASP and Ref.~\cite{Roising2019prr} TB sets, respectively, compared to $362$ meV in the parameter set used above. As a result, both the $B_{1g}$ and $A_{2g}$ pairing states are less favourable. However, the value of $(t_{1}-t_{2})$ for both of these is large enough such that the $A_{2g}$ pairing state can still be obtained if $x$ is large enough, but too small to make the $B_{1g}$ pairing state more competitive than the $A_{1g}$ pairing state. Along the $x$ axis of Fig.~\ref{fig1}, we find instead that for these two parameter sets there is a transition from $A_{1g}$ ($s$) $\rightarrow A_{1g}+iA_{2g}$ ($s+ig$) $\rightarrow A_{2g}$ ($g$). The transitions occur at $x\approx 0.43$ and $x\approx0.48$ for the VASP parameter set, and at $x\approx 0.6$ and $x\approx0.67$ using the TB set in Ref.~\cite{Roising2019prr}. Therefore, the $A_{2g}$ pairing state, which manifests with $g$-wave \textbf{k} dependence, is robust in the regime where the dominant energy scales are $\xi_{\textbf{k}}^{-}$ and $\lambda$, provided $\lambda_{B_{2g}}^{0}$ is a significant fraction of $\lambda$, with the precise value depending on the other kinetic terms.

\vspace{-3mm}
\section{\label{Five}Summary and Discussion}

We have shown that higher angular momentum pairings such as $d_{x^2-y^2}$- and $g_{xy(x^2-y^2)}$-wave arise from even-parity interorbital spin-triplet pairing in the presence of SOC. The local Hund's interaction provides an attractive channel for these pairing states within MF theory, yet they are stabilized by SOC and appear as \textbf{k}-dependent intraband pseudospin-singlet pairings. Pairing states in nontrivial irreps of the crystal point group can be found via the interplay between \textbf{k}-SOC and the dispersions of the orbitals involved. With three $t_{2g}$ orbitals, we find that when the difference between the orbital dispersions of the quasi-1D orbitals is large enough, the atomic SOC and SOC in the $B_{2g}$ channel alone are sufficient to give rise to $d$-wave and $g$-wave pairings, without requiring significant $B_{1g}$ or $A_{2g}$ forms of \textbf{k}-SOC.

We have applied this idea to Sr$_{2}$RuO$_{4}$ and find three multicomponent interorbital states in our analysis that break time-reversal symmetry, depending on the SOC: (i) $s+id_{x^2-y^2}$, (ii) $d_{x^2-y^2}+ig_{xy(x^2-y^2)}$, and (iii) $d_{xz}+id_{yz}$. Using the TB parameters obtained by first-principles calculations with atomic SOC and \textbf{k}-SOC, we find that both $A_{1g}$ and $B_{1g}$ components are nonzero, forming an $s+id_{x^2-y^2}$ state at $T=0$. We note that the presence of $s$ and $d_{x^2-y^2}$ pairings for small \textbf{k}-SOC is qualitatively consistent with the findings of other theoretical studies using spin-fluctuation mediated pairing without \textbf{k}-SOC \cite{Romer2019PRL,ZhangPRB2018}.

The SOC can be renormalized by electron-electron interactions, as shown for Sr$_{2}$RuO$_{4}$ \cite{zhang2016PRL,Kim2018PRL,Tamai2019PRX} as well as Sr$_{2}$RhO$_{4}$ \cite{LiuPRL2008}, and by a two-site exact diagonalization study which found an enhancement of the induced interatomic SOCs due to the Hund's coupling \cite{IsobePRB2014}. Thus, to understand the effects of SOC, we increase the \textbf{k}-SOC and find that for larger $|\lambda_{B_{2g}}^{0}|/\lambda$, both $B_{1g}$ and $A_{2g}$ states can be obtained, separated by a time-reversal symmetry breaking $B_{1g}+iA_{2g}$ pairing state ($d_{x^2-y^2}+ig_{xy(x^2-y^2)}$). While our calculation is at $T=0$, within the $d+ig$ region there is a point where the $d$- and $g$-wave gap components are equal, and therefore the transition temperatures of the two components are expected to coincide. This pairing state has been proposed previously from a phenomenological standpoint \cite{Kivelson2020npj, yuan2021} to account for a wide variety of the experimental results for Sr$_{2}$RuO$_{4}$. For instance, such a state is compatible with the observed discontinuity in the $c_{66}$ elastic modulus \cite{Ghosh2020,Benhabib2020},  time-reversal symmetry breaking \cite{Luke1998Nature,Xia2006prl, Kidwingira2006science}, vertical line nodes along the BZ diagonals \cite{DeguchiJPS2004,Hassinger2017PRX,sharma2020PNAS}, neutron and nuclear magnetic resonance Knight shift data \cite{Pustogow2019Nature,Ishida2019, Chronister2020, Steffens2019PRL}, and a splitting of time-reversal symmetry breaking and $T_{c}$ under uniaxial strain \cite{Grinenko2020}. We also note that contradictory evidence such as the absence of transition splitting in specific heat data under uniaxial strain \cite{Li_pnas2021}, and the observation that time-reversal symmetry breaking tracks $T_{c}$ under hydrostatic pressure \cite{grinenko2021unsplit} might be explained by a single-component $d_{x^2-y^2}$ or $g_{xy(x^2-y^2)}$ order parameter and spatially varying strain \cite{willaPRB2020,yuan2021}.


To pin down which pairing occurs in Sr$_{2}$RuO$_{4}$ within our scenario, it is necessary to determine the size of the \textbf{k}-SOC terms, as the higher angular momentum pairings, except $d$ in $s+id$, require them substantially larger than the DFT values. Our focus here is not to pin down the pairing of Sr$_{2}$RuO$_{4}$, but to demonstrate the mechanism for and competition among $d$, $g$, or $d+ig$ higher angular momentum pairings.  Quantifying the effects of correlations on the SOC strength is beyond the current paper.

Other open questions include both an analysis of odd-frequency pairings and their interplay with even-frequency interorbital spin-triplet states, and the effects of longer-range interactions. A recent study using such interactions has shown that nearly degenerate nodal $s$, $d_{xy}$, and helical spin-fluctuation mediated pairings are favored over $d_{x^2-y^2}$- and $g_{xy(x^2-y^2)}$-wave with only atomic SOC for sufficiently large nearest-neighbour repulsion \cite{romer2021}. The nodal $s+id_{xy}$ state proposed in Ref.~\cite{romer2021} is also consistent with a wide set of experimental data for Sr$_{2}$RuO$_{4}$ \cite{Luke1998Nature,Xia2006prl,Kidwingira2006science,Pustogow2019Nature,Ishida2019,Chronister2020,Steffens2019PRL,Ghosh2020,Benhabib2020,Grinenko2020}, and determination of the precise positions of the gap nodes in Sr$_{2}$RuO$_{4}$ will be key to differentiating this state from others such as a $d+ig$ state. As the situation regarding which of the many proposed states best fit the experimental data is far from settled, we have instead aimed to identify a mechanism for $g$-wave pairing in the present paper. Incorporating \textbf{k}-SOC and longer-range interactions within a realistic model is therefore an interesting problem left for future study. Finally, we would like to comment that our paper is applicable to other multiorbital materials with sizable Hund's and SOC.

\vspace{1em}
\begin{acknowledgments}
This work was supported by the Natural Sciences and Engineering Research Council of Canada (NSERC) Discovery Grant No. 2016-06089, the Center for Quantum Materials at
the University of Toronto, the Canadian Institute for
Advanced Research, and the Canada Research Chairs Program. Computations were performed on the
Niagara supercomputer at the SciNet HPC Consortium.
SciNet is funded by the Canada Foundation for Innovation under the auspices of Compute Canada, the Government of Ontario; Ontario Research Fund - Research Excellence, and the University of Toronto.
\end{acknowledgments}

\bibliography{main}

\section*{Appendix}

\renewcommand{\theequation}{A\arabic{equation}}
\setcounter{equation}{0}

\renewcommand{\thesubsection}{\arabic{subsection}}

\subsection{SOC in the \texorpdfstring{E\textsubscript{g}}{Eg} and \texorpdfstring{A\textsubscript{2g}}{A2g} channels}

The SOC in the $E_{g}$ channel discussed in the main text is
\begin{equation}    \label{egsoc}
\begin{aligned}
    H_{\text{SOC}}^{E_{g}} = &~8i\lambda_{E_{g}}^{0}\sum_{\textbf{k}\sigma\sigma'}\sigma_{\sigma\sigma'}^{z}\sin{\frac{k_x}{2}}\cos{\frac{k_y}{2}}\sin{\frac{k_z}{2}}c_{xz,\textbf{k}\sigma}^{\dagger}c_{xy,\textbf{k}\sigma'}\\
    &\hspace{-6mm}-8i\lambda_{E_{g}}^{0}\sum_{\textbf{k}\sigma\sigma'}\sigma_{\sigma\sigma'}^{z}\cos{\frac{k_x}{2}}\sin{\frac{k_y}{2}}\sin{\frac{k_z}{2}}c_{yz,\textbf{k}\sigma}^{\dagger}c_{xy,\textbf{k}\sigma'}\\
    &\hspace{-6mm}-8i\lambda_{E_{g},1d}^{0}\sum_{\textbf{k}\sigma\sigma'}\sigma^{x}_{\sigma\sigma'}\sin{\frac{k_x}{2}}\cos{\frac{k_y}{2}}\sin{\frac{k_z}{2}}c_{yz,\textbf{k}\sigma}^{\dagger}c_{xz,\textbf{k}\sigma'}
    \\&\hspace{-6mm}-8i\lambda_{E_{g},1d}^{0}\sum_{\textbf{k}\sigma\sigma'}\sigma^{y}_{\sigma\sigma'}\cos{\frac{k_x}{2}}\sin{\frac{k_y}{2}}\sin{\frac{k_z}{2}}c_{yz,\textbf{k}\sigma}^{\dagger}c_{xz,\textbf{k}\sigma'}  \\&\hspace{-6mm}+\text{H.c.},
\end{aligned}
\end{equation}
and the SOC in the $A_{2g}$ channel is
\begin{equation}
    \begin{aligned}
    H_{\text{SOC}}^{A_{2g}} = &-i\sum_{\textbf{k}\sigma\sigma'}\lambda_{\textbf{k}}^{A_{2g}}\sigma_{\sigma\sigma'}^{x}c_{yz,\textbf{k}\sigma}^{\dagger}c_{xy,\textbf{k}\sigma'}\\[6pt]&-i\sum_{\textbf{k}\sigma\sigma'}\lambda_{\textbf{k}}^{A_{2g}}\sigma_{\sigma\sigma'}^{y}c_{xz,\textbf{k}\sigma}^{\dagger}c_{xy,\textbf{k}\sigma'} + \text{H.c.},
    \end{aligned}
\end{equation}
where the form factor contains $g$-wave \textbf{k} dependence, $\lambda_{\textbf{k}}^{A_{2g}} = 8\lambda_{A_{2g}}^{0}\sin{k_x}\sin{k_y}(\cos{k_x}-\cos{k_y})$. We have neglected this SOC in the current paper as it originates from further-neighbour hopping processes and will be significantly smaller than all other SOC terms.

\subsection{Intraband pairing: \texorpdfstring{A\textsubscript{1g}}{A1g}, \texorpdfstring{B\textsubscript{1g}}{B1g}, and \texorpdfstring{B\textsubscript{2g}}{B2g} SOCs}

Here, we present the additional contributions to the intraband pairing neglected in Sec.~\hyperref[Three]{III} of the main text. Assuming every term in Eq.~(\ref{3orbMatrix}) is finite, i.e., incorporating the atomic SOC, $\lambda$, and the SOC with $B_{2g}$, $\lambda_{\textbf{k}}^{B_{2g}}$, and $B_{1g}$, $\lambda_{\textbf{k}}^{B_{1g}}$, form factors, the intraband pairing contains contributions in each of the 1D channels in Eqs.~(\ref{pairing_irreps}). Using the terms from Eqs.~(\ref{pairing_irreps}), we apply the transformation to the band basis given by Eq.~(\ref{last_trsf}). The resulting intraband pairing is
\begin{equation}
\begin{aligned}
    \widetilde{H}_{\text{pair}} = i&\Delta_{\textbf{k}}^{\beta*}(c_{\beta,\textbf{k}+}c_{\beta,-\textbf{k}-} -c_{\beta,\textbf{k}-}c_{\beta,-\textbf{k}+}) \\[6pt]&+i \Delta_{\textbf{k}}^{\gamma*} (c_{\gamma,\textbf{k}+}c_{\gamma,-\textbf{k}-} -c_{\gamma,\textbf{k}-}c_{\gamma,-\textbf{k}+}) + \text{H.c.},\\[6pt]
\end{aligned}
\end{equation}
where $\displaystyle \Delta_{\textbf{k}}^{\beta*} = \Delta_{\textbf{k},1}^{*} + \Delta_{\textbf{k},2}^{*}$ and $\displaystyle \Delta_{\textbf{k}}^{\gamma*} = -\Delta_{\textbf{k},1}^{*} + \Delta_{\textbf{k},2}^{*}$ and
\begin{subequations}
\label{full_gap}
    \begin{align}
    &\Delta_{\textbf{k},1}^{*} =~\frac{1}{E_{\widetilde{\beta},\textbf{k}}}\Big\{~ \Delta_{B_{2g}}^{*}\bigl[ \lambda^{B_{2g}}_{\textbf{k}} - \frac{2\lambda\bigl(t_{\textbf{k}} +\lambda^{B_{2g}}_{\textbf{k}}\bigr)}{E_{1d,\textbf{k}}} \bigr] \label{full_gap_1}
     \\[8pt]&\hspace{18mm}\notag+\Delta_{B_{1g}}^{*}\bigl[ \lambda^{B_{1g}}_{\textbf{k}} + \frac{\lambda\xi_{\textbf{k}}^{-} - 2\lambda\lambda^{B_{1g}}_{\textbf{k}}}{E_{1d,\textbf{k}}}\bigr] \\[8pt]&\hspace{18mm}\notag +\Delta_{A_{1g},1}^{*}\bigl[ \lambda + \frac{2\lambda^2 -2\lambda^{B_{2g}}_{\textbf{k}}t_{\textbf{k}} + \xi_{\textbf{k}}^{-}\lambda^{B_{1g}}_{\textbf{k}}}{E_{1d,\textbf{k}}} \bigr] \\[8pt] &\hspace{18mm}\notag + \Delta_{A_{1g},2}^{*}~\frac{\lambda}{E_{1d,\textbf{k}}}\widetilde{\xi}_{\textbf{k}}^{-}\\[8pt] &\hspace{18mm}\notag  +  \Delta_{A_{2g}}^{*}\frac{2\lambda^{B_{1g}}_{\textbf{k}}t_{\textbf{k}} + \xi_{\textbf{k}}^{-}\lambda^{B_{2g}}_{\textbf{k}}}{E_{1d,\textbf{k}}}~\Big\}, \\[8pt]
    &\Delta_{\textbf{k},2}^{*} = \Delta_{A_{1g},2}^{*}~\frac{\lambda}{E_{1d,\textbf{k}}}. \label{full_gap_2}
    \end{align}
\end{subequations}
The pairings, $\Delta_{\textbf{k}}^{\beta}$ and $\Delta_{\textbf{k}}^{\gamma}$ contain higher angular momentum contributions such as $d_{x^2-y^2}$- and $g_{xy(x^2-y^2)}$-wave via the interplay between SOC and kinetic energy. Considering the pairing in a two-band model with the $\alpha,\gamma$ bands instead by projecting out the $\widetilde{\beta}$ band, gives the same result, but with $E_{1d,\textbf{k}}\rightarrow - E_{1d,\textbf{k}}$. The $A_{1g}$, $B_{1g}$, and $B_{2g}$ channels all contain direct contributions from SOC in the associated channel. Additionally, other terms in each of these channels, as well as in the $A_{2g}$ channel, appear through products of SOCs in other channels with TB terms to give form factors with the proper symmetry. Due to the nature of the $d_{yz}$ and $d_{xz}$ orbitals, $\xi_{\textbf{k}}^{-}$ will generally provide the largest energy scale. If we further assume the next largest parameter to be the atomic SOC, $\lambda$, and that $\lambda_{B_{2g}}^{0}$ is the largest \textbf{k}-SOC parameter, the $B_{1g}$ and $A_{2g}$ pairing states can potentially both be competitive, depending on the interorbital hopping, $t_{\textbf{k}}$.


Taking the limit of purely atomic SOC, and assuming only the spin-triplet order parameters in the $A_{1g}$/$B_{1g}$ channels are nonzero as in Sec.~\hyperref[Three]{III}, the effective gaps reduce to
\begin{equation}
\begin{aligned}
    \Delta_{\textbf{k},1}^{*} = &\frac{1}{E_{\widetilde{\beta},\textbf{k}}}\Big\{~\Delta_{B_{1g}}^{*}\frac{\lambda\xi_{\textbf{k}}^{-}}{E_{1d,\textbf{k}}} \\[6pt]&+ \Delta_{A_{1g},1}^{*}\bigl[\lambda + \frac{2\lambda^2}{E_{1d,\textbf{k}}}\bigr] + \Delta_{A_{1g},2}^{*}\frac{\lambda}{E_{1d,\textbf{k}}} \widetilde{\xi}_{\textbf{k}}^{-}~\Big\},
\end{aligned}
\end{equation}
and $\Delta_{\textbf{k},2}^{*}$ remains as above. Therefore, in the limit that $\frac{|\xi_{\textbf{k}}^{-}|}{E_{1d,\textbf{k}}}\approx 1$ over most of the FS, the projection onto the intraband pairing in the $B_{1g}$ channel will be comparable to the $A_{1g,1}$ channel, to order $\lambda$. Note that both $A_{1g,1}$ and $A_{1g,2}$ channels are always finite in an $A_{1g}$ pairing state, therefore it is not expected that a $B_{1g}$ pairing would be stabilized over $A_{1g}$ with only atomic SOC. However, based on the full expression for the intraband pairing above, other terms such as the one proportional to $\lambda_{\textbf{k}}^{B_{2g}}t_{\textbf{k}}$ can favour the $B_{1g}$ pairing state by suppressing the gap in the $A_{1g}$ channel. For instance, when $t_{6}$ is the largest interorbital hopping parameter in $t_{\textbf{k}}$, the $A_{1g}$ pairing will be suppressed by the term proportional to $\lambda_{\textbf{k}}^{B_{2g}}t_{\textbf{k}} \sim \lambda_{B_{2g}}^{0}t_{6}(\sin{k_x}\sin{k_y})^2$, which appears with the opposite sign to $\lambda$ in the $A_{1g}$ pairing component.

Similarly, with only $B_{2g}$ SOC and the spin-triplet order parameters making up the $B_{2g}$/$A_{2g}$ channels, the effective gap is
\begin{equation}
\begin{aligned}
    \Delta_{\textbf{k},1}^{*} = \frac{1}{E_{\widetilde{\beta},\textbf{k}}}\Big\{~&\Delta_{B_{2g}}^{*}\lambda_{\textbf{k}}^{B_{2g}} + \Delta_{A_{2g}}^{*}\frac{\xi_{\textbf{k}}^{-}\lambda_{\textbf{k}}^{B_{2g}}}{E_{1d,\textbf{k}}}~\Big\}.
\end{aligned}
\end{equation}
Here, the two gap contributions can be approximately equal in the same limit as above. However, allowing for other kinetic terms and SOCs can lead to a larger gap in the $A_{2g}$ channel through both $\lambda$ and $t_{\textbf{k}}$.

\subsection{Intraband pairing: \texorpdfstring{E\textsubscript{g}}{Eg} SOC}
In this section, we examine the effects of the $E_{g}$ SOC and interorbital hoppings on the pairing, using a similar two-band analysis as above. As before, we first consider the mixing in the $(d_{yz},d_{xz})$ subspace through $t_{\textbf{k}}$ and $\lambda$, and then incorporate the coupling to the $d_{xy}$ orbital. We assume this occurs only through the $\sigma^{z}$ component of the $E_{g}$ SOC, given in Eq.~(\ref{egsoc}), as well as through the interorbital hoppings, $t_{\textbf{k}}^{yz/xy}$ and $t_{\textbf{k}}^{xz/xy}$ in Eqs.~(\ref{interorbhoppings}). Defining $\phi_{\textbf{k}+(-)}^{\dagger}=(c_{yz,\textbf{k}\uparrow(\downarrow)}^{\dagger},c_{xz,\textbf{k}\uparrow(\downarrow)}^{\dagger},c_{xy,\textbf{k}\uparrow(\downarrow)}^{\dagger})$, the three-orbital Hamiltonian is 
\begin{equation}
\begin{aligned}
&H_{0}+H_{\text{SOC}}=\sum_{\textbf{k},s=\pm}\phi_{\textbf{k}s}^{\dagger}B_{\textbf{k}s}\phi_{\textbf{k}s},\\[6pt]
&B_{\textbf{k}s} = 
\hspace{-1mm}
\setlength\arraycolsep{0.01mm}
 \begin{pmatrix} 
    \xi_{\textbf{k}}^{yz} & t_{\textbf{k}}+is\lambda & -is\lambda_{\textbf{k}}^{Y} + t_{\textbf{k}}^{yz/xy} \\[7pt]
    t_{\textbf{k}}-is\lambda & \xi_{\textbf{k}}^{xz} &is\lambda_{\textbf{k}}^{X} + t_{\textbf{k}}^{xz/xy} \\[7pt]
     is\lambda_{\textbf{k}}^{Y}+t_{\textbf{k}}^{yz/xy}& -is\lambda_{\textbf{k}}^{X} + t_{\textbf{k}}^{xz/xy} & \xi_{\textbf{k}}^{xy}
\end{pmatrix},
\end{aligned}
\end{equation}
where $\lambda_{\textbf{k}}^{Y} = 8\lambda_{E_{g}}^{0}\cos{\frac{k_x}{2}}\sin{\frac{k_y}{2}}\sin{\frac{k_z}{2}}$ and $\lambda_{\textbf{k}}^{X} = 8\lambda_{E_{g}}^{0}\sin{\frac{k_x}{2}}\cos{\frac{k_y}{2}}\sin{\frac{k_z}{2}}$. We transform the above three-orbital matrix to a two-band model by diagonalizing in the $(d_{yz},d_{xz})$ subspace yielding the two bands, $(\widetilde{\alpha},\widetilde{\beta})$, and project out the $\widetilde{\alpha}$ band. In the new basis, $\widetilde{\phi}_{\textbf{k}+(-)}^{\dagger}=(c_{\widetilde{\beta},\textbf{k}\uparrow(\downarrow)}^{\dagger}, c_{xy,\textbf{k}\uparrow(\downarrow)}^{\dagger})$, the Hamiltonian is
\begin{equation}
\begin{aligned}
&\widetilde{H}_{0}=\sum_{\textbf{k},s=\pm}\widetilde{\phi}_{\textbf{k}s}^{\dagger}\widetilde{B}_{\textbf{k}s}\widetilde{\phi}_{\textbf{k}s},\\[6pt]
&\widetilde{B}_{\textbf{k}s} = 
 \begin{pmatrix} 
    \xi_{\textbf{k}}^{\widetilde{\beta}} & \eta_{\widetilde{\beta},\textbf{k}}^{R}+is\eta_{\widetilde{\beta},\textbf{k}}^{I}  \\[7pt]
     \eta_{\widetilde{\beta},\textbf{k}}^{R}-is\eta_{\widetilde{\beta},\textbf{k}}^{I} & \xi_{\textbf{k}}^{xy}
\end{pmatrix},
\end{aligned}
\end{equation}
where the hybridization between $\widetilde{\beta}$ and $d_{xy}$ is given by
\begin{equation}
\begin{aligned}
\eta_{\widetilde{\beta},\textbf{k}}^{R} &= \frac{|\tilde{f}_{\textbf{k}}|}{\sqrt{t_{\textbf{k}}^2+\lambda^2}}\bigl(\lambda\lambda_{\textbf{k}}^{Y}-t_{\textbf{k}}t_{\textbf{k}}^{yz/xy}\bigr)-|\tilde{g}_{\textbf{k}}|t_{\textbf{k}}^{xz/xy} ,\\
\eta_{\widetilde{\beta},\textbf{k}}^{I} &= \frac{|\tilde{f}_{\textbf{k}}|}{\sqrt{t_{\textbf{k}}^2+\lambda^2}}\bigl(t_{\textbf{k}}\lambda_{\textbf{k}}^{Y} + \lambda t_{\textbf{k}}^{yz/xy}\bigr) - |\tilde{g}_{\textbf{k}}|\lambda_{\textbf{k}}^{X}.
\end{aligned}
\end{equation}
Note that in this case, each term contains $d_{xz}$- or $d_{yz}$-wave \textbf{k} dependence reflecting the $E_{g}$ symmetry.

We diagonalize the above Hamiltonian by
\begin{equation} \label{appendix_trsf}
\begin{aligned}
\begin{pmatrix}
   c_{\widetilde{\beta},\textbf{k}\sigma} \\[6pt]
   c^{xy}_{\textbf{k}\sigma} \\
 \end{pmatrix}\hspace{-1.5mm}
 \setlength\arraycolsep{0.01mm}
 = \hspace{-1.5mm}\begin{pmatrix} 
    \frac{\eta_{\sigma}+1}{2}f_{\textbf{k}}-\frac{\eta_{\sigma}-1}{2}f_{\textbf{k}}^{*} & -g_{\textbf{k}}  \\[6pt]
  g_{\textbf{k}} & \frac{\eta_{\sigma}+1}{2}f_{\textbf{k}}^{*}-\frac{\eta_{\sigma}-1}{2}f_{\textbf{k}}
\end{pmatrix}\hspace{-2.5mm}
\begin{pmatrix}
   c_{\beta,\textbf{k}s} \\[6pt]
   c_{\gamma,\textbf{k}s} \\
 \end{pmatrix},
\end{aligned}
\end{equation}
where the coefficients are the same as those given in the main text, but with $\eta_{\widetilde{\beta},\textbf{k}}$ defined above. Since we have only included the $z$ component of the $E_{g}$ SOC, the spin-triplet order parameters that contribute to the intraband pairing are $d_{xz/xy}^{z}$ and $d_{xy/yz}^{z}$. The resulting pairing is
\begin{equation}
\begin{aligned}
    \widetilde{H}_{\text{pair}} = i&\Delta_{\textbf{k}}^{*}\bigl[(c_{\beta,\textbf{k}+}c_{\beta,-\textbf{k}-} -c_{\beta,\textbf{k}-}c_{\beta,-\textbf{k}+}) \\[6pt]&- (c_{\gamma,\textbf{k}+}c_{\gamma,-\textbf{k}-} -c_{\gamma,\textbf{k}-}c_{\gamma,-\textbf{k}+}) + \text{H.c.},\\[6pt]
\end{aligned}
\end{equation}
where
\begin{equation}
\begin{aligned}
\Delta_{\textbf{k}}^{*} =\frac{-1}{E_{\widetilde{\beta},\textbf{k}}}\Big\{&d_{xy/yz}^{z*}\bigl[~\lambda_{\textbf{k}}^{Y}(1+\frac{\xi_{\textbf{k}}^{-}}{E_{1d,\textbf{k}}})\\[6pt]&\hspace{-2mm}-\frac{2}{E_{1d,\textbf{k}}}(\lambda_{\textbf{k}}^{X}t_{\textbf{k}} + \lambda t_{\textbf{k}}^{xz/xy})\bigr] \\[6pt]&\hspace{-3.5mm}+d_{xz/xy}^{z*}\bigl[~\lambda_{\textbf{k}}^{X}(1-\frac{\xi_{\textbf{k}}^{-}}{E_{1d,\textbf{k}}})\\[6pt]&\hspace{-2mm}-\frac{2}{E_{1d,\textbf{k}}}(\lambda_{\textbf{k}}^{Y}t_{\textbf{k}}+ \lambda t_{\textbf{k}}^{yz/xy})~\bigr]\Big\}.
\end{aligned}    
\end{equation}
The pairing contribution proportional to $d_{xy/yz}^{z*}$ with $d_{yz}$ symmetry is related to that proportional to $d_{xz/xy}^{z*}$ with $d_{xz}$ symmetry under a $C_{4}$ rotation, reflecting the overall $E_{g}$ symmetry of the pairing. Also, due to the $d_{xz}/d_{yz}$ \textbf{k} dependence of the interorbital hopping between the quasi-1D and $d_{xy}$ orbitals, there is a contribution to the $E_{g}$ pairing through the terms proportional to $\lambda t_{\textbf{k}}^{xz/xy}$ and $\lambda t_{\textbf{k}}^{yz/xy}$.

\end{document}